\documentclass[aps,prc,superscriptaddress,twocolumn,showpacs,letter]{revtex4-1}  
\usepackage{graphicx}  
\usepackage{bm}        
\usepackage{amssymb}   
\usepackage{amsmath}
\usepackage{longtable}
\usepackage{dcolumn}
\usepackage{multirow}
\usepackage{verbatim}

\begin{document}

\title{A Search for $0^+$ States in $^{50}$Cr: Implications for the Superallowed $\beta$-decay of $^{50}$Mn}
\author{K.G.~Leach}\email{kleach@mines.edu}\address{Department of Physics, Colorado School of Mines, Golden, CO 80401, USA}\address{TRIUMF, 4004 Wesbrook Mall, Vancouver, British Columbia V6T 2A3, Canada}
\author{P.E.~Garrett}\affiliation{Department of Physics, University of Guelph, Guelph, Ontario N1G 2W1, Canada}
\author{G.C.~Ball}\affiliation{TRIUMF, 4004 Wesbrook Mall, Vancouver, British Columbia V6T 2A3, Canada}
\author{P.C.~Bender}\altaffiliation{Present address: National Superconducting Cyclotron Laboratory, Michigan State University, East Lansing, Michigan 48824, USA}\affiliation{TRIUMF, 4004 Wesbrook Mall, Vancouver, British Columbia V6T 2A3, Canada}
\author{V.~Bildstein}\affiliation{Department of Physics, University of Guelph, Guelph, Ontario N1G 2W1, Canada}
\author{B.A.~Brown}\affiliation{National Superconducting Cyclotron Laboratory, Michigan State University, East Lansing, Michigan 48824, USA}\affiliation{Department of Physics and Astronomy, Michigan State University, East Lansing, Michigan 48824, USA}
\author{C.~Burbadge}\affiliation{Department of Physics, University of Guelph, Guelph, Ontario N1G 2W1, Canada}
\author{T.~Faestermann}\affiliation{Physik Department, Technische Universit\"at M\"unchen, D-85748 Garching, Germany}
\author{B.~Hadinia}\affiliation{Department of Physics, University of Guelph, Guelph, Ontario N1G 2W1, Canada}
\author{J.D.~Holt}\affiliation{TRIUMF, 4004 Wesbrook Mall, Vancouver, British Columbia V6T 2A3, Canada}
\author{A.T.~Laffoley}\altaffiliation{Present address: Grand Acc\'el\'erateur National d'Ions Lourds (GANIL), CEA/DSM-CNRS/IN2P3, Boulevard Henri Becquerel, 14076 Caen, France}\affiliation{Department of Physics, University of Guelph, Guelph, Ontario N1G 2W1, Canada}
\author{D.S.~Jamieson}\altaffiliation{Present address: Department of Physics and Astronomy, Stony Brook University, Stony Brook, NY 11794-3800}\affiliation{Department of Physics, University of Guelph, Guelph, Ontario N1G 2W1, Canada}
\author{B.~Jigmeddorj}\affiliation{Department of Physics, University of Guelph, Guelph, Ontario N1G 2W1, Canada}
\author{A.J.~Radich}\affiliation{Department of Physics, University of Guelph, Guelph, Ontario N1G 2W1, Canada}
\author{E.T.~Rand}\altaffiliation{Present Address: AECL Chalk River Laboratories, 286 Plant Rd. Stn 508A, Chalk River, Ontario K0J 1J0, Canada}\affiliation{Department of Physics, University of Guelph, Guelph, Ontario N1G 2W1, Canada}
\author{S.R.~Stroberg}\affiliation{TRIUMF, 4004 Wesbrook Mall, Vancouver, British Columbia V6T 2A3, Canada}
\author{C.E.~Svensson}\affiliation{Department of Physics, University of Guelph, Guelph, Ontario N1G 2W1, Canada}
\author{I.S.~Towner}\affiliation{Cyclotron Institute, Texas A\&M University, College Station, Texas 77843-3366, USA}
\author{H.-F.~Wirth}\affiliation{Fakult\"at f\"ur Physik, Ludwig-Maximilians-Universit\"at M\"unchen, D-85748 Garching, Germany}

\date{\today}

\begin{abstract}
A $^{52}$Cr$(p,t)$$^{50}$Cr two-neutron pickup reaction was performed using the Q3D magnetic spectrograph at the Maier-Leibnitz-Laboratorium in Garching, Germany.  Excited states in $^{50}$Cr were observed up to an excitation energy of 5.3~MeV.  Despite significantly increased sensitivity and resolution over previous work, no evidence of the previously assigned first excited $0^+$ state was found.  As a result, the $0^+_2$ state is reassigned at an excitation energy of $E_x=3895.0(5)$~keV in $^{50}$Cr.  This reassignment directly impacts direct searches for a non-analogue Fermi $\beta^+$ decay branch in $^{50}$Mn.  These results also show better systematic agreement with the theoretical predictions for the $0^+$ state spectrum in $^{50}$Cr using the same formalism as the isospin-symmetry-breaking correction calculations for superallowed nuclei.  The experimental data are also compared to {\it ab-initio} shell-model predictions using the IM-SRG formalism based on $NN$ and $3N$ forces from chiral-EFT in the $pf$-shell for the first time.
\end{abstract}

\maketitle
Tests of the Standard Model through precision measurements of nuclear decay properties have proven to be a valuable tool in experimental subatomic physics~\cite{Nav13}.  Of these investigations, superallowed Fermi nuclear $\beta$ decay data are among the most important, as they currently provide the most precise determination of the vector coupling strength in the weak interaction, $G_V$~\cite{Har15}.  This is uniquely possible in this class of radioactive decay since the transition operator that connects the initial and final $0^+$ states is (to first order) independent of the axial-vector contribution to the weak interaction.  By obtaining $G_V$, the up-down element of the Cabibbo-Kobayashi-Maskawa (CKM) quark-mixing matrix, $V_{\mathrm{ud}}$ is also extracted from the precise $0^+\rightarrow0^+$ nuclear $\beta$ decay $ft$ values.  The $ft$ values are experimentally determined through measurements of three quantities; the half-life, the decay $Q$-value, and the branching fraction of the superallowed decay mode~\cite{Har15}.
\newline\indent
In order to obtain the level of precision required for Standard Model tests, corrections to the experimental $ft$-values must also be made to obtain nucleus-independent ${\cal F}t$ values,
\begin{eqnarray}
{\cal F}t\equiv ft(1+\delta_R)(1-\delta_C)=\frac{2\pi^3\hbar^7\ell n(2)}{2G_V^2m_e^5c^4(1+\Delta_R)},
\label{Ft_value}
\end{eqnarray}
where $\delta_R$ is a transition-dependent radiative correction, $\Delta_R$ is a transition-independent radiative correction, and $\delta_C$ is a nucleus-dependent isospin-symmetry-breaking (ISB) correction.  Although relatively small ($\sim1\%$), these corrections are crucial due to the very precise ($\leq0.1\%$) $ft$-values~\cite{Har15} that result from decades of high-precision measurements at both stable and radioactive ion-beam facilities~\cite{Har15}.  In fact, the current uncertainty on $G_V$, and consequently $V_{ud}$, is dominated by the precision of these theoretical corrections.  With a value of 2.361(38)\%~\cite{Mar06}, the largest fractional uncertainty of any individual correction term is due to the transition-independent radiative correction, $\Delta_R$.  Despite the large uncertainty, the QED formalism that is used in the calculation of this quantity is well understood, suggesting that the central value is accurate.  This is not as clear for the ISB corrections~\cite{Sat09}, which have a similarly large uncertainty contribution in the extraction of $G_V$, but require very difficult nuclear-structure calculations in relatively large model spaces~\cite{Tow10}.
\newline\indent
The current extraction of $G_V$ and $V_{ud}$ from the superallowed data uses the shell-model-calculated ISB corrections of Towner and Hardy (TH), which have been the benchmark for nearly forty years.  In that time, the experimental $ft$ values have become increasingly more precise - particularly in the last decade - and as such, the model-space truncations~\cite{Tow10} and small deficiencies that exist in this formalism~\cite{Mil08} must be investigated.  Additionally, with increasing computational power, more exact theoretical treatments which were out of reach in the past are now under investigation~\cite{Cal09,Sat09,Lia09,Aue09,Rod12,Sat16}, thus providing useful insight into where some of the older phenomenological approaches may be incomplete~\cite{Tow10b}.  Despite these new techniques, the TH Woods-Saxon shell-model approach remains the standard today due to the high level of experimental testing~\cite{Bha08,Mel11} and guidance~\cite{Mol15,Lea13a,Lea13b} the formalism has been exposed to.  However, due to the dramatic implications of a deviation from unity in the top-row sum of the CKM matrix resulting from a shift in the $\delta_C$ calculations, continued experimental verification and testing of these quantities is critical.
\newline\indent
In the TH theoretical framework, a separation of $\delta_C$ into a sum of two terms is performed.  The first, $\delta_{C1}$, results from different configuration mixing between the parent and daughter nuclear states in the superallowed decay.  The second, $\delta_{C2}$, accounts for the imperfect radial overlap between the initial and final nuclear wave functions.  If the shell-model effective interaction was truly isospin invariant, the parent and daughter analogue-state wave functions would be identical and would lead to all non-analogue $\beta$ decay transitions to $0^+$ states in the daughter nucleus being strictly forbidden~\cite{Tow10}.
\newline\indent
For the calculation of $\delta_{C1}$ in particular, a relatively small model space is used (typically one major oscillator shell)~\cite{Tow15}.  In this model space, the charge-dependent terms are added to the charge-independent effective Hamiltonian of the shell-model.  These additional terms are separately adjusted to reproduce the $b$ and $c$ coefficients of the isobaric multiplet mass equation (IMME) for each $T=1$ $J^\pi=0^+$ superallowed $\beta$ transition.  For all of the $0^+$ states in a given model space with the same total isospin, the effect of this mixing is to deplete the analogue transition strength by the sum of the mixing into all of the non-analogue states,
\begin{eqnarray}
\delta_{C1}\approx\sum_n\delta_{C1}^n,
\label{deltac1_sum}
\end{eqnarray}
where $n$ is a counting index for excited $0^+$ states.  In most cases, the bulk of the mixing is associated with the first few excited non-analogue $0^+$ states~\cite{Tow10}.  Since the calculations of $\delta_{C1}$ are highly sensitive to the specific model that is used, the TH approach adopts two strategies to reduce model dependencies.  The first method is to scale the $\delta_{C1}$ values by the square of the ratio between the theoretical and experimentally observed excited $0^+$-state energies,
\begin{eqnarray}
\frac{(\Delta E)^2_{SM}}{(\Delta E)^2_{exp.}}.
\end{eqnarray}
The second method was originally reported in the formalism of Ormand and Brown~\cite{Orm85}, where the charge-dependent part of the effective interaction is constrained such that the coefficients of the IMME~\cite{Mac14} are properly reproduced.  It is the former of these two methods that is of particular interest to the work presented here, since it requires accurate experimental information regarding excited $0^+$ state energies.
\newline\indent
Motivated by recent half-life and Gamow-Teller $\beta$-decay branching ratio measurements by the RISING collaboration at GSI~\cite{Mol15}, the theoretical ISB corrections for the $A=42,46,50$, and 54 superallowed systems were recently re-visited in Ref.~\cite{Tow15}.  For the case of $^{50}$Mn specifically, the parameters of the IMME had significantly changed since the previous calculations reported in Ref.~\cite{Har15}, which resulted in a reduction of the $^{50}$Mn $\delta_{C1}$ value by roughly a quarter of its central value from 0.045(20)\% to 0.035(20)\%~\cite{Tow15}.  This reduction may be consistent with previous measurements in other superallowed systems~\cite{Hag94,Pie03,Fin08,Lea08,Dun13}, which have suggested that the non-analogue $\beta$-decay strength is over-predicted by a significant amount.  Therefore, it is important to provide experimental bench-marking of these calculations, especially as they are updated and refined.
\newline\indent
As mentioned above, if isospin were an exact symmetry, the parent and daughter analogue-state wave functions would be identical in superallowed Fermi $\beta$ decay.  Since isospin is not an exact symmetry, configuration mixing generates non-analogue Fermi $\beta$-decay branches which allows for the direct measurement of the $\delta_{C1}^n$ from the observed $\beta$-decay branching ratios, $B_n$, to the $n$ excited $0^+$ states~\cite{Hag94},
\begin{eqnarray}
\delta_{C1}^n\approx\left(\frac{f_0}{f_n}\right)B_n\frac{(1-\delta_{C1})}{B_0}\approx\left(\frac{f_0}{f_n}\right)B_n,
\label{deltac1}
\end{eqnarray}
where $f_0$ and $f_n$ are the phase-space integrals for decay to the ground state and $n^{\mathrm{th}}$ excited $0^+$ state, respectively.
\newline\indent
The only direct experimental search for non-analogue $0^+\rightarrow0^+$ Fermi $\beta$ decay of $^{50}$Mn is reported in Ref.~\cite{Hag94}.  In that Letter, no definitive non-analogue decay branch was observed, although two new Gamow-Teller (GT) $\beta$-decay branches to $1^+$ states in $^{50}$Cr were measured, with branching fractions of 570~ppm and 6.8~ppm to states at 3628 and 4998~keV, respectively.  This resulted in a quoted superallowed branching ratio of 99.9423(30)\%~\cite{Hag94}.  The authors noted, however, that the information on excited $0^+$ states in $^{50}$Cr ``was very imprecise", and that ``the evidence for five possible $0^+$ states below 5~MeV were not well established".  The imprecise knowledge of the $0^+$ state locations in the daughter nucleus thus limited their ability to place limits on possible non-analogue transitions.  In particular, the $\beta$-delayed $\gamma$ transition from the first excited $0^+$ state in $^{50}$Cr is expected to carry the majority of the non-analogue strength~\cite{Tow15}.  Unfortunately, in the 22 years that have passed since that $\beta$-decay work was published, the situation regarding knowledge of these $0^+$ states has not changed.  Due to the strong recent motivation for a direct search in $^{50}$Mn for non-analogue $\beta$ decay branches at radioactive ion-beam (RIB) facilities, the present work reports a high-sensitivity search for $0^+$ states in $^{50}$Cr using the $(p,t)$ two-neutron-transfer reaction.
\newline\indent
\begin{figure*}[t!]
\rotatebox{-90}{\includegraphics[width=0.4\linewidth]{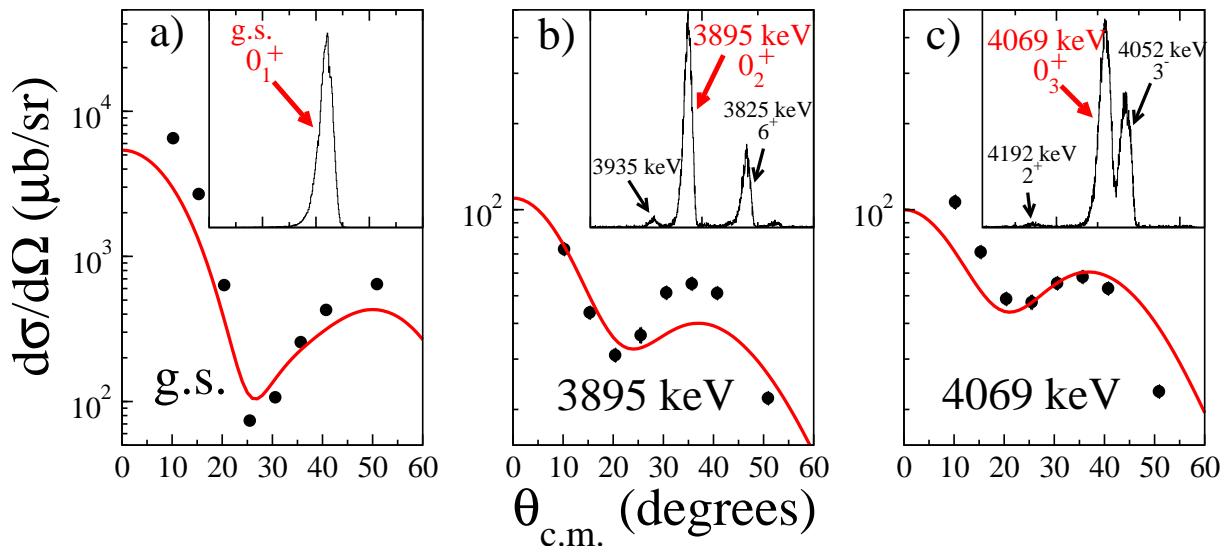}}
\caption{(Color online) A comparison of the \textsc{fresco} calculated $L=0$ angular distributions (red curve) using shell-model generated two-nucleon amplitude files, to the experimental data for a) the ground-state, and excited states at b) 3895.0(5)~keV and c) 4068.8(5)~keV. Experimental errors are shown, however in most cases they are smaller than the data points.  The curves are normalized to the data to provide a better shape comparison for the identification of the $0^+$ states.  The inset in each panel shows the triton energy spectrum at $10^\circ$ for transfer to the respective states.}
\label{DWBA_to_data}
\end{figure*}
The experiment was performed at the Maier-Leibnitz-Laboratorium (MLL) of Ludwig-Maximilians-Universit\"at (LMU) and Technische Universit\"at M\"unchen (TUM) in Garching, Germany.  A $\sim1~\mu$A beam of protons was accelerated to 24~MeV using the MP tandem Van de Graaff accelerator, and was incident on a $>99\%$ isotopically pure, $\sim130$~$\mu$g/cm$^2$ $^{52}$Cr target with a $\sim10~\mu$g/cm$^2$ carbon backing.  The reaction products were momentum analyzed using a Q3D magnetic spectrograph, and the resulting particles were detected at the focal plane using a cathode-strip detector~\cite{Wir00} with a full-width at half-maximum (FWHM) energy resolution of roughly 10~keV.  Outgoing tritons were observed at eight angles between $10^\circ$ and $50^\circ$, up to an excitation energy in $^{50}$Cr of 5.4~MeV using seven different momentum settings of the spectrograph.  These settings were altered slightly at various angles to remove peaks from reactions on light-mass target impurities in an effort to reduce system dead time.  A $0^{\circ}$ Faraday cup inside the target chamber was used to determine the number of beam particles incident on the target by integrating the total current.  Using this information, cross-sections were determined and angular distributions were extracted.  Due to the extremely large negative reaction $Q$-value for $^{52}$Cr$(p,t)$, no stable target was suitable for energy calibration.  Instead, outgoing triton energies were benchmarked to $^{50}$Cr excitation energies using well-known states from the evaluated data~\cite{ENSDF}.  To further improve the accuracy and precision of the calibration, the triton energy dependence as a function of position from this work and Refs.~\cite{Lea13a,Lea13b} were combined to extrapolate the fit parameters above $\sim3$~MeV.  This parameter-extrapolation, along with overlapping position spectra, allowed for an energy calibration of the higher-energy states.
\begin{table}[b!]
\centering
\caption{A comparison of the shell-model predicted $L=0$ transfer cross-sections to the experimental data at $10^\circ$ for the three observed $0^+$ states in this work.  Cross-sections are given as d$\sigma$/d$\Omega$ in $(\mu$b/sr).}
\label{tab:compare}
\begin{tabular}{c|ccc}
\hline\hline
 & g.s. & 3.895~MeV & 4.069~MeV \\
\hline
Experiment & 6500(300) & 72(3) & 106(6) \\
Theory & 1150 & 64 & 5 \\
\hline\hline
\end{tabular}
\end{table}
\newline\indent
The identification of the $0^+$ states was conducted through a comparison of the extracted angular-distribution data to two-nucleon-transfer reaction calculations performed using the coupled-channel, finite-range DWBA software \textsc{fresco}~\cite{FRESCO}.  The \textsc{fresco} calculations use shell-model final-state wave functions from two-nucleon-amplitude (tna) files, and include both simultaneous and sequential transfer of the two neutrons.  The tna files were generated from shell-model calculations using the GXPF1A interaction~\cite{Hon05}.  The global proton and triton optical-model parameters of Varner \textit{et al.}~\cite{Var91} and Li, Liang, and Cai~\cite{Li07} were used, respectively.  The proton parameters remained constant for all calculations, with the triton values calculated as a function of outgoing triton energy.  To provide a test of the shell-model theory used, the \textsc{fresco} calculated cross-sections for the observed $0^+$ states in Fig.~\ref{DWBA_to_data} are compared to the experimental data at $10^\circ$ in Table~\ref{tab:compare}.  In most cases, the primary feature used for identification of the observed $0^+$ states was a characteristic increasing cross-section at forward angles, as shown in Fig.~\ref{DWBA_to_data}.  However, as the outgoing triton energies decreased in energy (corresponding to higher excitation energies in $^{50}$Cr), the DWBA-calculated angular distributions for $L=0$ transfer show more subtle features due to the increased sequential-transfer component, thus making the identification of $0^+$ states more difficult.
\newline\indent
\begin{figure}[t!]
\rotatebox{-90}{\includegraphics[width=0.75\linewidth]{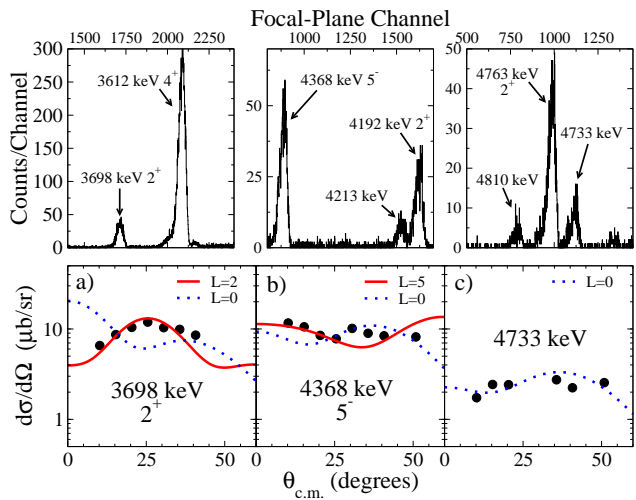}}
\caption{(Color online) Experimental angular distributions for the observed levels at a) 3697.9(6)~keV, b) 4368(3)~keV, and c) 4733(5)~keV, shown with DWBA predictions for $L=0$ transfers (blue dashed line) and the $L$ transfer recommended from this work (red line).  Experimental errors are shown, however in most cases they are smaller than the data points.  Above each angular distribution panel is the triton energy spectrum at $\theta_{\rm Q3D}=10^\circ$ for the region of interest.}
\label{Other_Plots}
\end{figure}
The compiled nuclear data in Ref.~\cite{ENSDF} reports five excited $0^+$ states in $^{50}$Cr from the three previous $^{52}$Cr$(p,t)^{50}$Cr experiments~\cite{Bae71,Ram72,Yam82}, two of which were performed more than 40 years ago.  With significantly improved resolution and sensitivity compared to these previous measurements, only two excited $0^+$ states in $^{50}$Cr were definitively observed in the work presented here: 3895.0(5)~keV and 4068.8(5)~keV, while a third excited $0^+$ state at 4733(5) keV is also plausible.  The agreement (or disagreement) with the evaluated $0^+$-state data in Ref.~\cite{ENSDF} is discussed on a level-by-level basis below.
\newline\indent{\bf 3694~keV state:}  This state was first reported in Ref.~\cite{Ram72} using a 31.4~MeV $^{52}$Cr$(p,t)^{50}$Cr reaction, and has not been confirmed since, including the more recent $(p,t)$ measurement in Ref.~\cite{Bae71}.  The work presented here observes only one state in this region, which is consistent with an evaluated $2^+$ state from Ref.~\cite{ENSDF} at 3697.9(6)~keV, shown in Fig.~\ref{Other_Plots} a).  Given the level of sensitivity and high final-state energy resolution of the present work, with no evidence for an additional state nearby, the first excited $0^+$ state in $^{50}$Cr is reassigned below.
\newline\indent{\bf 3850~keV state:}  Ref.~\cite{Bae71} reports a possible $0^+$ state through the observation of a doublet at 3850(20)~keV using a 27~MeV $^{52}$Cr$(p,t)^{50}$Cr reaction.  Although the quoted energy is not in $1\sigma$ agreement with the 3895.0(5)~keV state reported here, due to the relatively thick target used in that work ($\sim1$~mg/cm$^2$), a final-state energy resolution of roughly 90~keV was reported, which likely encapsulated the energy region of interest.  A state at 3898(7)~keV was also observed in Ref.~\cite{Ram72} with a characteristic increasing cross-section at low angles, but a $J^\pi$ was not assigned.  This state was also assigned as a $0^+$ in a $^{48}$Ti$(^3$He$,n)^{50}$Cr reaction in Ref.~\cite{Boh75}.  The present work therefore confirms these three observations with a definite assignment of $J^\pi=0^+$ and an excitation energy of 3895.0(5)~keV.
\newline\indent{\bf 4050~keV state:}  The only previous observation of this state is reported in Ref.~\cite{Bae71} as a doublet of states at 4050~keV with a proposed $0^+$ state contribution.  With the significantly increased energy resolution achieved in the work presented here, this doublet of $3^-$ and $0^+$ states was resolved, and is shown in Fig.~\ref{DWBA_to_data}~c).  It is therefore assigned as the second excited $0^+$ state in $^{50}$Cr at an excitation energy of 4068.8(5)~keV.
\newline\indent{\bf 4350~keV state:}  The only observation of this state is reported in Ref.~\cite{Bae71} with a very weak population and very large uncertainties on the reported cross-sections.  This state overlaps in energy with the first $5^-$ state at 4367.0(4)~keV reported in the evaluated data~\cite{ENSDF}.  A peak at 4368(3)~keV observed in this work is in excellent energy agreement with this reported $5^-$ state, and demonstrates relatively good agreement with the $L=5$ transfer shown in Fig.~\ref{Other_Plots} b).  The \textsc{fresco} calculated $L=0$ curve for the shell-model-predicted third excited $0^+$ state shows a very flat angular distribution which cannot be immediately dismissed.  However, given the large amount of experimental data in Ref.~\cite{ENSDF} to support $J^\pi=5^-$, the present work adopts the same assignment for this state.
\newline\indent{\bf 4728~keV state:}
The final $0^+$ state below 5~MeV listed in Ref.~\cite{ENSDF} is at an energy of 4728(7)~keV.  The previous $(p,t)$ work of Ref.~\cite{Bae71} reported this state at an energy of 4750~keV with a $\sim10~\mu$b/sr cross-section.  In the present work, two states were observed in this general energy region at 4733(5)~keV and 4763(5)~keV.  The second state at 4763(5)~keV is in good energy and $L$-transfer agreement with the evaluated $2^+$ state at 4772(7)~keV in Ref.~\cite{ENSDF}.  The 4733(5)~keV state is in energy agreement with the evaluated data for the $0^+$, however the cross-section magnitude and angular distribution shown in Fig.~\ref{Other_Plots} c) are so featureless that no definitive conclusion can be made.  Therefore, the previous assignment is neither supported, nor refuted by the work presented here.
\newline\indent
\begin{figure}[t!]
\rotatebox{-90}{\includegraphics[width=0.5\linewidth]{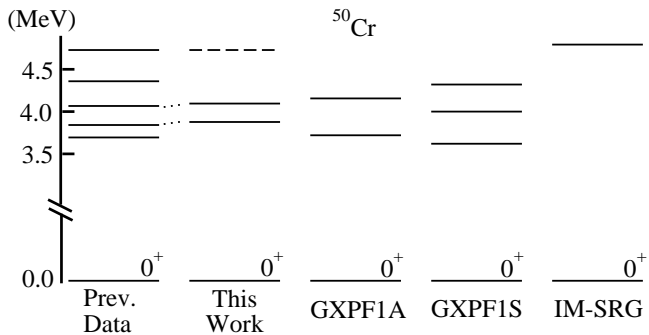}}
\caption{A comparison of the $0^+$ state excitation energies in $^{50}$Cr from (a) the evaluated data in Ref.~\cite{ENSDF}, (b) this work, and shell-model calculations using (c) the GXPF1A interaction~\cite{Hon05}, (d) a modified GXPF1 interaction~\cite{Hon02} (see text), and (e) {\it ab-initio} IM-SRG~\cite{Str16a} (see text).}
\label{zeroplus_states}
\end{figure}
A comparison of the experimental spectrum of $0^+$ state energies in $^{50}$Cr to three theoretical models, shown in Fig.~\ref{zeroplus_states}, also supports the observations presented here.  In particular,  shell-model calculations using the GXPF1A interaction~\cite{Hon05} predicts two $0^+$ states that are very close in energy to experiment.  The agreement with a modified (older version) of the GXPF1 interaction~\cite{Hon02} (labelled here as GXPF1S) is somewhat worse, with three excited $0^+$ states predicted below 4.5~MeV. The modified GXPF1 interaction is the current method employed for calculating the wavefunctions used in the TH $\delta_{C1}$ determination.
\newline\indent
The experimental data are also compared to {\it ab-initio} in-medium similarity renormalization group (IM-SRG) calculations~\cite{Her16} based on two- $(NN)$ and three-nucleon $(3N)$ forces from chiral Effective Field Theory (EFT)~\cite{Epe09,Mac11}.  The $NN$ and $3N$ forces of Refs.~\cite{Ent03} and~\cite{Nav07}, respectively, are evolved with the free-space SRG~\cite{Bog09} to the resolution scale $\lambda_{\mathrm{SRG}}=1.88~\mathrm{fm}^{-1}$ and transformed to the Hartree-Fock basis with model-space truncations of $2n+\ell\leq e_{\mathrm{max}}=12$ and $e_1+e_2+e_3\leq E_{\mathrm{3max}}=14$.  In the standard approach, the Magnus formulation of the IM-SRG~\cite{Mor15} is used to decouple consistently both the core energy and valence-space Hamiltonian of interest~\cite{Tsu12}, with $3N$ forces normal ordered with respect to the valence-space core~\cite{Bog14,Cac15}.  More recently to account for $3N$ forces between valence nucleons~\cite{Cae14}, a targeted normal ordering scheme was developed in which $3N$ forces are normal ordered with respect to the nearest subshell closure~\cite{Str16a}.  In this work an improved procedure is adopted, where ensemble normal ordering accurately includes $3N$ forces specifically in the nucleus of interest~\cite{Str16b}. For $^{50}$Cr the $pf$ shell outside a $^{40}$Ca core is taken for both protons and neutrons, and the valence-space diagonalization is performed with the NuShellX shell model code~\cite{Bro14}.
\newline\indent
\begin{figure}[t!]
\begin{center}\rotatebox{-90}{\includegraphics[width=0.8\linewidth]{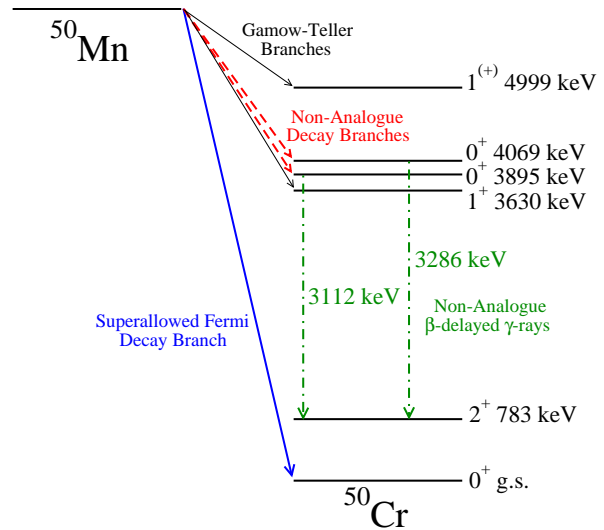}}\end{center}
\caption{(Color online) A partial schematic of the $\beta^+/$EC decay scheme for $^{50}$Mn.  The two lowest-lying $0^+$ states in $^{50}$Cr presented in this work indicate that non-analogue Fermi $\beta^+$ decay branches (red dashed) to these states would emit characteristic $E2$ $\gamma$ rays at $E_\gamma=3111.7(5)$ and 3285.5(5)~keV (green dot-dashed).}
\label{DecaySchemes}
\end{figure} 
Although the IM-SRG prediction of the $0_2^+$, shown in Fig.~\ref{zeroplus_states}, is too high compared with experiment and phenomenology, the fact that it agrees to the level of several hundred keV is notable since it represents the first such calculation with no local fitting to data, as well as the first application of valence-space IM-SRG to the $pf$ shell. It should also be noted that the input chiral Hamiltonians, when used in a number of many-body methods, are insufficient to describe nuclei in the calcium region and heavier~\cite{Bin14}.  Thus, forces with improved saturation properties are currently being explored~\cite{Eks15,Sim16}.  With the rapid progression of {\it ab-initio} methods in the medium-mass region, it is likely that these theoretical approaches may soon be able to calculate the superallowed ISB corrections from first principles.
\newline\indent
The reassignment of the first excited $0^+$ state has two important consequences for the TH $\delta_{C1}$ calculations.  The first is a small effect, and is related to the energy scaling for the $\delta_{C1}$ term in $^{50}$Mn, which is now lowered from $\delta_{C1}=0.035(20)\%$ quoted in Ref.~\cite{Tow15} to $\delta_{C1}=0.031(20)\%$.  The second consequence has much larger implications, as it directly impacts searches for a non-analogue decay branch in $^{50}$Mn at RIB facilities, such as with the GRIFFIN spectrometer~\cite{Sve14} at TRIUMF.  Since the first excited $0^+$ states is expected to have the strongest isospin mixing with the isobaric analogue state~\cite{Tow15}, it would therefore contain the majority of the non-analogue $\beta$ decay strength from $^{50}$Mn.  The reassignment of the $0^+$ states therefore suggests that current work towards identifying (or placing limits) on any non-analogue decay strength in this system should focus efforts on observing a 3111.7(5)~keV $\gamma$ ray that would result from the $0^+_2\rightarrow2^+_1$ transition to the 783.32(3)~keV state in $^{50}$Cr (Fig.~\ref{DecaySchemes}).  A direct observation of any non-analogue decay strength in these systems will continue to provide important benchmarks for the various calculation methods used in the extraction of $G_V$ from the superallowed Fermi $\beta$-decay data.
\newline\indent
To summarize, a 24~MeV $^{52}$Cr$(p,t)$$^{50}$Cr experiment was performed to search for excited $0^+$ states in the superallowed $\beta$-decay daughter nucleus of $^{50}$Mn.  Two excited $0^+$ states in $^{50}$Cr were identified at excitation energies of 3895.0(5)~keV and 4068.8(5)~keV, and a third excited $0^+$ at 4733(5)~keV remains tentatively assigned.  No evidence for two of the five previously reported $0^+$ states in the evaluated data were found.  As a result, the first excited $0^+$ state in $^{50}$Cr is reassigned at an energy of 3895.0(5)~keV.  These results show better agreement with the $0^+$-state spectrum predicted by phenomenological shell-model calculations.  Ongoing searches for a non-analogue branch in the superallowed $\beta$ decay of $^{50}$Mn should therefore look for the characteristic 3111.7(5)~keV $\gamma$ ray which would result from a de-excitation of the first excited $0^+$ state in $^{50}$Cr.
\newline\indent
This work was supported in part by the Natural Sciences and Engineering Research Council of Canada (NSERC), the Ontario Ministry of Economic Development and Innovation, and NSF grant PHY-1068217.  C.E.S. acknowledges support from the Canada Research Chair (CRC) program.  The authors thank the MLL operation staff for their efforts in providing a high-quality beam of 24~MeV protons for the duration of the experiment.

\end{document}